\documentclass[pre,twocolumn,epsfig,epsf]{revtex4-2}
\usepackage{%refcheck,
amsmath,amsthm,amsfonts,amscd,mathrsfs,pifont,bm}
\usepackage{eucal,graphicx,color}
\usepackage{esint}

\makeatletter
\@for\next:={int,iint,iiint,iiiint,dotsint,oint,oiint,sqint,sqiint,
  ointctrclockwise,ointclockwise,varointclockwise,varointctrclockwise,
  fint,varoiint,landupint,landdownint}\do{%
    \expandafter\edef\csname\next\endcsname{%
      \noexpand\DOTSI
      \expandafter\noexpand\csname\next op\endcsname
      \noexpand\ilimits@
    }%
  }
\makeatother

\begin{document}

\title  {Power spectra and autocovariances of level spacings beyond the Dyson conjecture}

\author {Roman Riser$^{1,2}$, Peng Tian$^{1}$, and Eugene Kanzieper$^{1,3}$}

\affiliation
       {
       $^1$~Department of Mathematics, Holon Institute of Technology, Holon 5810201, Israel\\
       $^2$~Department of Physics and Research Center for Theoretical Physics and Astrophysics, University of Haifa, Haifa 3498838, Israel \\
       $^3$~Department of Physics of Complex Systems, Weizmann Institute of Science, Rehovot 7610001, Israel \\
       \\
       \rm{(Submitted 23 January 2023; accepted 2 March 2023; published 17 March 2023)}
       }

\begin  {abstract}
Introduced in the early days of random matrix theory, the autocovariances $\delta I^j_k={\rm cov}(s_j, s_{j+k})$ of level spacings $\{s_j\}$ accommodate a detailed information on correlations between individual eigenlevels. It was first conjectured by Dyson that the autocovariances of distant eigenlevels in the unfolded spectra of infinite-dimensional random matrices should exhibit a power-law decay $\delta I^j_k\approx -1/\beta\pi^2k^2$, where $\beta$ is the symmetry index. In this Letter, we establish an exact link between the autocovariances of level spacings and their power spectrum, and show that, for $\beta=2$, the latter admits a representation in terms of a fifth Painlev\'e transcendent. This result is further exploited to determine an asymptotic expansion for autocovariances that reproduces the Dyson formula as well as provides the subleading corrections to it. High-precision numerical simulations lend independent support to our results. \\
\newline
Published in: \href{https://doi.org/10.1103/PhysRevE.107.L032201}{\textcolor{blue}{Phys. Rev. E {\bf 107}, L032201 (2023)}}
\end{abstract}

\maketitle
%\newpage\noindent
\textit{\textbf{Introduction.}}---Universal aspects of spectral fluctuations in generic quantum systems which are fully chaotic in the classical limit are accurately described by the random matrix theory~\cite{M-2004,PF-book} (RMT). This statement, known as the Bohigas-Giannoni-Schmit (BGS) conjecture~\cite{BGS-1984}, has summarized earlier attempts~\cite{MK-1979,CVG-1980,B-1981,Z-1981} ``to put in close contact two areas -- random matrix physics and the study of chaotic motion -- that have remained disconnected'' until the mid-eighties of the past century. Supported by a vast amount of experimental and numerical evidence, the emergence of universal statistical laws (which, in appropriate energy or time domains, are governed by global symmetries rather than by system specialties) has later been advocated within a field-theoretic~\cite{SUSY-1996-1997} and a semiclassical approach~\cite{Semiclassics-2002}.

To probe energy level fluctuations in the {\it unfolded} spectra of bounded quantum systems, a number of spectral statistical measures have been devised~\cite{M-2004}. While it is customary to distinguish between short- and long-range statistical indicators (highlighting spectral correlations on the local and global energy scales, respectively), we find it more appropriate -- in the context of this Letter -- to assign them to two alternative classes of (i) {\it ordinary} and (ii) {\it ordered} level statistics. The two clearly differ from each other on a formal level due to much different mathematical structures lurking behind them.

(i) The {\it ordinary} (linear) spectral statistics~\cite{Product-remark} describes energy spectrum as a whole without referring to a specific eigenlevel. In the random matrix theory setting, it deals with the fluctuation properties of a random variable
\begin{eqnarray}\label{reg-stat}
    X_N({\bm \lambda}) = \sum_{\ell=1}^N f_N(\lambda_\ell),
\end{eqnarray}
where $f_N(\lambda)$ is a (not necessarily linear) function of interest and
${{\bm \lambda}=\{\lambda_1,\dots,\lambda_N\}}$ are (possibly unfolded) eigenvalues of an $N\times N$ random matrix. Clearly, $X_N({\bm \lambda})$ is invariant under the ordering of $N$ eigenlevels. Typical representatives of the ordinary statistical indicators include two-point (or higher-order) correlation functions, or their integrated counterparts -- variance of the (fluctuating) number ${\mathcal N}(L)$ of eigenlevels in the interval of length $L$ or its higher-order cumulants.

(ii) The {\it ordered} level statistics refers to {\it individual} eigenlevels and thus cannot be defined without their ordering. (For one, any statistics dealing with level spacings is, unavoidably, ordered.) Generically, it can be formulated in terms of a random variable
\begin{eqnarray}\label{ord-stat}
    X_N({\bm \lambda};{\bm c}) = \sum_{\ell=1}^N c_\ell f_N(\lambda_\ell),
\end{eqnarray}
where (unfolded) random eigenvalues are ordered, that is ${\{\lambda_1 \le \dots \le \lambda_N\}}$, and a sequence of weights ${{\bm c}=\{c_1,\dots,c_N\}}$ is not a constant one, ${\bm c}\neq c \openone_N$. The latter is a game changer as it makes the moments of $X_N({\bm \lambda};{\bm c})$ depend on spectral correlation functions of {\it all} orders -- the feature which is not necessarily present in the ordinary level statistics. Taken together with the correlated nature of the RMT eigenvalues, this explains why the ordered level statistics remains a relatively unexplored territory.

From now on, we switch from the finite-$N$ spectra to the unfolded spectra of infinite-dimensional random matrices, described by the universal ${\rm Sine}_\beta$ point process~\cite{D-1962-CUE,S-2000,MNN-2019}. Such a setting, implied by the BGS conjecture, provides an effective calculational environment for quantifying universal spectral fluctuations in fully chaotic quantum systems.

Arguably, the most thoroughly studied~\cite{JMMS-1980,TW-1993,M-2004} example of the ordered statistics is the distribution of level spacing between consecutive eigenlevels. For the $\beta=2$ Dyson's symmetry class~\cite{M-2004}, associated with quantum chaotic systems with broken time-reversal symmetry, the level spacing distribution equals
\begin{eqnarray}\label{LSD-PV}
   P_\infty(s) = \frac{d^2}{ds^2} \exp \left(
            \int_{0}^{2 \pi s} \frac{\sigma_0(t;\zeta=1)}{t} dt
        \right),
\end{eqnarray}
where $\sigma_0(t;\zeta=1)$ is the {\it single member} ($\zeta=1$) of a family of one-parameter solutions to the fifth Painlev\'e equation
\begin{equation} \label{PV-eq-0}
    (t \sigma_0^{\prime\prime})^2 + (t\sigma_0^\prime -\sigma_0) \left[
            t\sigma_0^\prime -\sigma_0 + 4 (\sigma_0^\prime)^2
        \right]= 0,
\end{equation}
which are analytic at $t=0$ and satisfy the $\zeta$-dependent boundary condition
\begin{eqnarray}\label{bc-zero-0}
\!\!\!\sigma_0(t; \zeta)=  -\frac{t}{2\pi}\zeta -\left(\frac{t}{2\pi}\right)^2\zeta^2 + {\mathcal O}(t^3) \quad {\rm as} \quad t\rightarrow 0.
\end{eqnarray}
The same {\it family} of one-parameter solutions will later surface in a nonperturbative description of the autocovariances of level spacings which are the main focus of our study.

\textit{\textbf{From distribution of level spacings to their correlations.}}---Stunning in its appearance, the exact result Eq.~(\ref{LSD-PV}) is completely {\it local} as it provides no information about {\it correlations} between {\it different} spacings. To probe the latter, it is beneficial to define yet another ordered spectral statistics -- the autocovariances
\begin{eqnarray}\label{auto-cov}
    \delta I^j_k = {\rm cov}(s_j,s_{j+k}) = \langle s_j s_{j+k} \rangle -1
\end{eqnarray}
of level spacings located $|k|$ eigenlevels apart. Here, the $\ell$-th ($\ell \in {\mathbb{N}}$) level spacing~\cite{Rem-conventions} $s_\ell = \lambda_{\ell}-\lambda_{\ell-1}$ is associated with a sequence of ordered, {\it unfolded} eigenlevels ${\{0 \le \lambda_1 \le  \lambda_2 \le \dots \}}$; angular brackets $\langle \dots\rangle$ denote an appropriate ensemble averaging.

Little is known, let alone rigorously proven, about level spacing correlations. Two properties of autocovariances of level spacings are self-evident though: (i) describing the spectral bulk, $\delta I^j_k$ does not depend on the position $j$ of a reference eigenlevel being a function of $|k|$ only~\cite{M-2004} (for this reason we shall write $\delta I_{k}$ or $\delta I_{|k|}$ from now on); (ii) ${\delta I_{k} \rightarrow 0}$ as $k\rightarrow \infty$ since the correlations between spacings of extremely distant eigenlevels should eventually die out.

{\it How fast do the correlations weaken as the distance between eigenlevels grows?}  In the numerical study of nontrivial zeros of the Riemann zeta function, Odlyzko~\cite{O-1987} has quoted an unpublished conjecture by Dyson
\begin{eqnarray} \label{Dyson-conjecture}
    \delta I_{k} \approx \delta I_{k}^D = - \frac{1}{2\pi^2 k^2}
\end{eqnarray}
assumed to hold asymptotically for sufficiently large $k$. A heuristic argument in favor of Dyson's conjecture was outlined in Refs.~\cite{FMP-1978,BFFMPW-1981}, where it was argued that the variance ${\rm var}(\lambda_k)$ of the $k$-th ordered eigenlevel $\lambda_k$ in the unfolded spectrum can be related to the variance $\Sigma^2(k)$ of number of eigenlevels ${\mathcal N}(k)$ in the interval of {\it integer} length $k\gg 1$
\begin{eqnarray} \label{french-relation}
    {\rm var} (\lambda_k) \approx \Sigma^2(k) - \frac{1}{6}.
\end{eqnarray}
Detailed knowledge of the number variance~\cite{M-2004,PF-book}, combined with the equality~\cite{ROK-2020,Rem-conventions}
\begin{eqnarray}\label{theta-k-relation}
    \delta I_{k} = \frac{1}{2} \left[
        {\rm var}(\lambda_{k+1}) - 2 {\rm var}(\lambda_k) + {\rm var}(\lambda_{k-1})
    \right],
\end{eqnarray}
had produced~\cite{FMP-1978,BFFMPW-1981} the Dyson formula Eq.~(\ref{Dyson-conjecture}).

The same relation Eq.~(\ref{french-relation}) has later been employed by Bohigas and collaborators~\cite{BLS-1999,BLS-2001} in an attempt to study the autocovariance of level spacings {\it beyond} Dyson's conjecture. Two alternative approaches~\cite{BLS-1999,BLS-2001} brought out two different results for $\delta I_{|k|}$ thus questioning the validity of both. This discrepancy adds up to a somewhat obscure status~\cite{French-remark} of the asymptotic formula Eqs.~(\ref{french-relation}). Since the number variance $\Sigma^2(k)$ in its r.h.s. is a two-point spectral statistics whilst the variance ${\rm var}(\lambda_k)$ in the l.h.s. is clearly not, Eq.~(\ref{french-relation}) misses all the information about higher-order spectral correlations. How important is this lack of knowledge for the autocovariances?

\textit{\textbf{Motivation.}}---The scarce and inconclusive understanding of level spacings correlations in the random matrix theory summarized in the two last paragraphs suggests that a {\it nonperturbative} theoretical framework is required to tackle the problem.

In this Letter, we outline how such a framework can be built by linking the autocovariances of level spacings to yet another (ordered) spectral statistics known as the power spectrum~\cite{Rem-spacings-eigenlevels} of level spacings~\cite{O-1987} and the power spectrum of eigenlevels~\cite{RGMRF-2002,Rem-PS-DT}. To the best of our knowledge, their connection to the autocovariances has never been discussed in the literature.

Over the past two decades, the power spectra have emerged as an effective tool for studying both system-specific and universal properties of quantum systems. In particular, the power spectrum analysis reveals~\cite{RGMRF-2002} whether the corresponding classical dynamics is regular or chaotic, or a mixture~\cite{GRRFSVR-2005} of both, and encodes a ``degree of chaoticity''~\cite{PRPAG-2018}. In combination with other long- and short-range spectral fluctuation measures, the power spectrum statistics provides an effective way to identify system symmetries, determine a degree of incompleteness of measured spectra, and get the clues about systems internal dynamics. On the experimental side, the power spectrum was measured in Sinai \cite{FKMMRR-2006} and perturbed rectangular \cite{BYBLDS-2016b} microwave billiards, microwave networks \cite{BYBLDS-2016} and three-dimensional microwave cavities \cite{LBYBS-2018}. More recently, the power spectra surfaced in the studies~\cite{XR-2023} of the non-ergodic extended regime in physical and random matrix models.

\begin{figure}[t]
\includegraphics[width=0.45\textwidth]{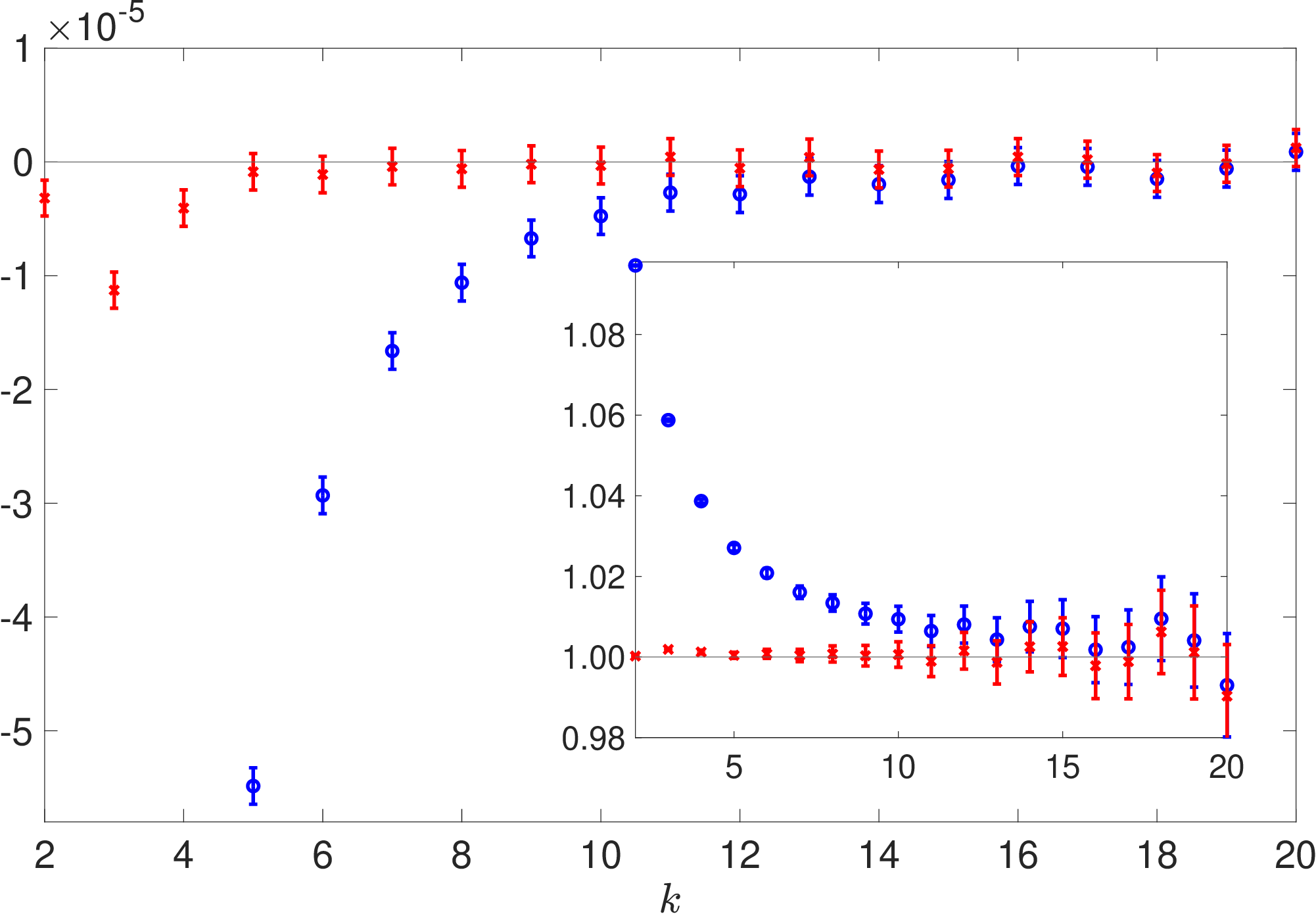}
\caption{Visualization of numerical data produced from $M=10^8$ samples of ${\rm CUE}(N)$
with $N=1024$. Blue circles and red crosses represent the $k$-dependence of the differences $\delta \mathcal{I}_k(N,M)-\delta I_k^D$ (blue) and
$\delta \mathcal{I}_k(N,M)-(\delta I_k^D+\delta I_k^S)$ (red) between numerically computed autocovariances $\delta \mathcal{I}_k(N,M)$ and either the leading order Dyson's term [Eq.~(\ref{Dyson-conjecture})] or the improved analytical prediction [Eq.~(\ref{main-02})]. The vertical bars around the data points are
$99\%$ confidence intervals. The inset shows the ratios $\delta \mathcal{I}_k(N,M)/\delta I_k^D$ and ${\delta \mathcal{I}_k(N,M)/(\delta I_k^D+\delta I_k^S)}$, respectively, for the same numerical data. A detailed description of the numerical procedure is given at the end of the Letter.}
\label{Fig-1}
\end{figure}

This Letter brings another dimension to the power spectrum statistics by positioning it as an elaborative theoretical tool tailor-made for a detailed analysis of correlations between level spacings which refused a nonperturbative treatment for several decades.

\textit{\textbf{Main results and discussion.}}---Our approach rests on the {\it exact} relation ($\beta=2$)
\begin{eqnarray}\label{main-01}
    S^{{\rm sp}}_\infty(\omega) = \sum_{k\in{\mathbb Z}} \delta I_{|k|} z^k \qquad \qquad\qquad \qquad\qquad \qquad \qquad \nonumber\\
    \quad = \frac{1}{\pi} {\rm Re} \int_{0}^{\infty} d\lambda \exp \left(
        \int_{0}^{\lambda} \frac{dt}{t} \sigma_0(t; \zeta = 1 - e^{i \omega})
    \right)\quad
\end{eqnarray}
between the autocovariances $\delta I_{|k|}$ of level spacings and their power spectrum $S^{\rm sp}_\infty (\omega)$~~\cite{Rem-spacings-eigenlevels}, the latter being equivalently defined by Eqs.~(\ref{PS-spacings}) and (\ref{PS-levels-stationary}) below. Here, $z=e^{i\omega}$ whilst $\sigma_0(t; \zeta=1-z)$ belongs to a one-parameter family of the fifth Painlev\'e transcendent defined by Eqs.~(\ref{PV-eq-0}) and (\ref{bc-zero-0}). Equation (\ref{main-01}) is the {\it first main result} of this Letter. Notice, that contrary to the level spacing distribution Eq.~(\ref{LSD-PV}), described in terms of the fifth Painlev\'e transcendent $\sigma_0(t;\zeta)$ at $\zeta=1$, a more detailed knowledge of $\sigma_0(t;\zeta)$ -- on the entire unit circle $|1-\zeta|=1$ -- is needed to account for {\it correlations} of level spacings.

Viewed as a Fourier series, Eq.~(\ref{main-01}) can be inverted to bring a nonperturbative representation of autocovariances $\delta I_{|k|}$ for all $k \in {\mathbb Z}$ [see Eq.~(\ref{inv-exact}) below]. Such an exact representation is potentially useful for both numerical and analytical analysis of $\delta I_{|k|}$. Leaving the former, computationally demanding route for a future study, we shall opt for the latter, aiming to generate an asymptotic expansion for autocovariances of level spacings as $k\rightarrow\infty$. In particular, we derive the first three terms of such an expansion,
\begin{eqnarray}\label{main-02}
    \delta I_{k} = &-&\frac{1}{2\pi^2 k^2} \nonumber\\
    &-&\frac{3}{2\pi^4 k^4} \left(\log (2\pi k) + \gamma -\frac{11}{6}\right)+ o(k^{-4}), \quad{}
\end{eqnarray}
where $\gamma=0.5772...$ is the Euler constant. While the leading-order term in Eq.~(\ref{main-02}) merely validates the Dyson conjecture [Eq.~(\ref{Dyson-conjecture})], the subleading $k^{-4}\log k$ and $k^{-4}$ terms (to be denoted $\delta I_k^S$) are completely new. Equation~(\ref{main-02}) is the {\it second main result} of this Letter.

In Fig.~\ref{Fig-1}, we confront the analytical prediction Eq.~(\ref{main-02}) with the autocovariances of level spacings computed numerically for large-dimensional Haar distributed unitary matrices. Even though one cannot expect that the asymptotic result Eq.~(\ref{main-02}) will provide a fair approximation for level spacing autocovariances at very low values of $k$, the red-marked data clearly indicate that our formula for $\delta I_k^D+\delta I_k^S$ fits a high-precision numerics very well starting with $k = 5$. Indeed, for all $k\ge 5$, the horizontal zero line -- corresponding to a virtual situation where numerically evaluated autocovariances would coincide exactly (for all $k$) with the analytical prediction $\delta I_k^D + \delta I_k^S$ -- lies inside the red-marked confidence intervals. In sharp contrast, comparison of numerical data with the Dyson formula Eq.~(\ref{Dyson-conjecture}) reveals a disagreement up to $k=14$ as the blue-marked confidence bars start to repeatedly hit the zero line only afterwards.

\textit{\textbf{Derivation of the first result.}}---We prove Eq.~(\ref{main-01}) in three steps. (i) First, we define the power spectrum of level {\it spacings}~\cite{O-1987}
as the limit
\begin{subequations} \label{PS-spacings}
\begin{eqnarray}  \label{PS-spacings-a}
    S^{\rm sp}_\infty (\omega) &=& \lim_{n \rightarrow \infty} S^{\rm sp}_n (\omega), \\
    \label{PS-spacings-b}
    S^{\rm sp}_n (\omega) &=&\frac{1}{n} \sum_{\ell,m=1}^{n}  {\rm cov}(s_\ell, s_m) z^{\ell-m}
\end{eqnarray}
\end{subequations}
[see notation specified below Eq.~(\ref{auto-cov})]. Here, $z=e^{i\omega}$ and $0 \le \omega \le \pi$. Stationarity of level spacings, supplemented by a sufficiently fast decay of autocovariances ${\delta I_k={\mathcal O}(k^{-p})}$ with $p>1$ as $k\rightarrow \infty$, ensures that the limit in the r.h.s. exists and approaches the continuous function
\begin{eqnarray}\label{PS-levels-stationary}
    S^{\rm sp}_\infty (\omega) = \sum_{k \in{\mathbb Z}} \delta I_{|k|}\,z^{k}.
\end{eqnarray}
(ii) Second, we define the power spectrum of {\it eigenlevels}~~\cite{Rem-spacings-eigenlevels,RGMRF-2002,ROK-2017}
\begin{subequations} \label{PS-eigenlevels}
\begin{eqnarray} \label{PS-eigenlevels-a}
    S^{\rm eig}_\infty (\omega) &=& \lim_{n \rightarrow \infty} S^{\rm eig}_n (\omega), \\
    \label{PS-eigenlevels-b}
    S^{\rm eig}_n (\omega) &=& \frac{1}{n} \sum_{\ell,m=1}^{n}  {\rm cov}(\lambda_\ell, \lambda_m) z^{\ell-m}
\end{eqnarray}
\end{subequations}
[see notation below Eq.~(\ref{auto-cov})], and further claim the identity
\begin{eqnarray}\label{eig-sp-link}
    S_\infty^{\rm eig}(\omega) = \frac{1}{4 \sin^2(\omega/2)} \big[ S_\infty^{\rm sp}(\omega) + S_\infty^{\rm sp}(0)\big],
\end{eqnarray}
which links the power spectrum of eigenlevels $S_\infty^{\rm eig}(\omega)$ to the power spectrum of spacings $S_\infty^{\rm sp}(\omega)$ for $\omega \in (0,\pi]$. In the particular case of the ${\rm Sine}_\beta$ point process, the contribution $S_\infty^{\rm sp}(0)$ nullifies due to the sum rule~\cite{P-1986} $\sum_{k\in {\mathbb Z}} \delta I_{|k|}=0$ which accounts for zero level compressibility. (iii) Third, the power spectrum of eigenlevels $S_\infty^{\rm eig}(\omega)$ for the ${\rm Sine}_2$ {determinantal point process (which is of our main interest) is known in terms of a fifth Painlev\'e transcendent [Eqs.~(\ref{PV-eq-0}) and (\ref{bc-zero-0})] owing to the recent study~\cite{RK-2023}. Combining Eqs.~(\ref{PS-levels-stationary}) and (\ref{eig-sp-link}) with Theorem~1.2 of Ref.~\cite{RK-2023}, which states that the power spectrum of {\it eigenlevels} equals
\begin{eqnarray}\label{Th12-quotation}
    S_\infty^{\rm eig}(\omega) &=& \frac{1}{4\pi \sin^2(\omega/2)} \nonumber\\
    &\times&
    {\rm Re} \int_{0}^{\infty} d\lambda \exp \left(
        \int_{0}^{\lambda} \frac{dt}{t} \sigma_0(t; \zeta = 1 - e^{i \omega})
    \right)\qquad
\end{eqnarray}
for all $\omega\in (0,\pi)$, we reproduce the announced Eq.~(\ref{main-01}).

It remains to justify the identity Eq.~(\ref{eig-sp-link}) which holds generically for random spectra with stationary level spacings. To proceed, we start with the finite-$n$ power spectrum of eigenlevels $S_n^{{\rm eig}}(\omega)$ defined by Eq.~(\ref{PS-eigenlevels-b}). As soon as ${\rm cov}(\lambda_\ell,\lambda_m) = \sum_{i=1}^{\ell}\sum_{j=1}^{m} {\rm cov}(s_i,s_j)$, one observes the relation
\begin{eqnarray} \label{link-finite-n}
    S_n^{{\rm eig}}(\omega) = \frac{1}{4\sin^2(\omega/2)}\!\! \left(\! S_n^{{\rm sp}}(\omega) + S_n^{{\rm sp}}(0)
    - \frac{2}{n} r_n(\omega)\! \right)\!, \qquad
\end{eqnarray}
where $S_n^{{\rm sp}}(\omega)$ is the finite-$n$ power spectrum of spacings, see Eq.~(\ref{PS-spacings-b}), while
\begin{eqnarray} \label{reminder-term}
    r_n(\omega) = {\rm Re\,}\sum_{\ell,m=1}^{n} \delta I_{|\ell-m|} z^{\ell-(n-1)}.
\end{eqnarray}
As $n\rightarrow \infty$, the first two terms in Eq.~(\ref{link-finite-n}) approach those in Eq.~(\ref{eig-sp-link}). The third term is of the order ${\mathcal O}(n^{-1})$ since the fast decay of autocovariances, assumed below Eq.~(\ref{PS-spacings}), keeps $r_n(\omega)$ bounded for any fixed $\omega \in (0,\pi]$. Hence, it can safely be dropped.

\textit{\textbf{Derivation of the second result.}}---To prove Eq.~(\ref{main-02}), we start with the Fourier inversion formula
\begin{eqnarray} \label{inv-exact}
    \delta I_{k} = \frac{1}{\pi} \int_{0}^{\pi} S_\infty^{\rm sp}(\omega)\, \cos(\omega k) d\omega
\end{eqnarray}
and calculate its asymptotics for $k \in {\mathbb N}$ large enough by making use of the stationary phase approximation. As ${k\rightarrow \infty}$, the integral Eq.~(\ref{inv-exact}) is dominated by vicinities of the endpoints. Their contributions can be determined by repeatedly employing integration by parts. In the vicinity of $\omega=0$, the required information will be extracted out of the small-$\omega$ expansion of the power spectrum $(\omega>0)$
\begin{eqnarray}\label{ps-small-omega}
    S_\infty^{\rm sp}(\omega) = \frac{\omega}{2\pi} + \frac{\omega^3}{4\pi^3} \log\left( \frac{\omega}{2\pi} \right) + {\mathcal O}(\omega^4)
\end{eqnarray}
which follows from Eq.~(\ref{eig-sp-link}) combined with Proposition~4.10 of Ref.~\cite{RK-2023}, see also Theorem~2.11 of Ref.~\cite{ROK-2020}. As for the upper integration bound, $2\pi$ periodicity and the evenness of $S_\infty^{\rm sp}(\omega)$ imply that its derivatives of odd orders with respect to $\omega$ vanish at $\omega=\pi$, that is ${S_\infty^{\rm sp}}^{(2j-1)}(\pi)=0$ for $j=1, 2, \dots$.

To achieve the accuracy $o(k^{-4})$ in the asymptotic expansion of $\delta I_k$, four integrations by parts are required. Spotting that both the first and second derivatives of the power spectrum $S_\infty^{\rm sp}(\omega)$ stay finite at $\omega=0$ whilst the third derivative possesses there a logarithmic singularity, we obtain after three consecutive integrations by parts:
\begin{eqnarray} \label{3-int-parts}
    \delta I_{k} = - \frac{1}{2\pi^2 k^2} &+& \frac{3}{2\pi^4 k^3}  \int_{0}^{\pi} \log\omega \,\sin(\omega k) d\omega \nonumber\\
     &+& \frac{1}{\pi k^3} \int_{0}^{\pi} \widetilde{{S_\infty^{\rm sp}}}^{(3)}(\omega)\,\sin(\omega k) d\omega,
\end{eqnarray}
where
\begin{eqnarray} \label{S-sp-omega-3}
    \widetilde{{S_\infty^{\rm sp}}}^{(3)}(\omega) = {S_\infty^{\rm sp}}^{(3)}(\omega) - \frac{3}{2\pi^3} \log\omega
\end{eqnarray}
is bounded for $\omega \in [0,\pi]$. Calculating the first integral in Eq.~(\ref{3-int-parts}) exactly while handling the second integral by parts, we derive:
\begin{eqnarray} \label{I-k-exact}
\delta I_k = - \frac{1}{2\pi^2 k^2} &-& \frac{3}{2\pi^4 k^4}\left( \log(2\pi k) - {\rm Ci}(\pi k) +\gamma -\frac{11}{6}\right) \nonumber\\
    &+& \frac{1}{\pi k^4} \int_{0}^{\pi} \widetilde{{S_\infty^{\rm sp}}}^{(4)}(\omega)
            \,\cos(\omega k) d\omega.
\end{eqnarray}
Here, ${\rm Ci}(x)$ is the cosine integral.

Two remarks are in order. (i) First, since $\widetilde{{S_\infty^{\rm sp}}}^{(4)}(\omega)$ is integrable on $\omega\in [0,\pi]$, the integral in Eq.~(\ref{I-k-exact}) is of the order $o(1)$, so that its contribution to $\delta I_k$ is of the order $o(k^{-4})$ as $k\rightarrow \infty$. (ii) Second, ${\rm Ci}(\pi k) = {\mathcal O}(k^{-2})$ provided $k\in {\mathbb N}$. Applying (i) and (ii) to Eq.~(\ref{I-k-exact}), we reproduce the announced asymptotic expansion Eq.~(\ref{main-02}).

\textit{\textbf{Description of numerical procedure.}}---To evaluate the level autocovariances numerically, we create ${M=10^8}$ samples of random ${\rm CUE}(N)$ spectra of the size ${N=1024}$. Denoting $\{ \theta_\ell^{(\alpha)}\}_{\ell=1}^N$ a set of ordered ${\rm CUE}(N)$ eigenangles generated in $\alpha$-th realization (${1\le \alpha \le M}$), we construct a set $\{ \tilde{s}_\ell^{(\alpha)} = \theta_{\ell+1}^{(\alpha)}- \theta_\ell^{(\alpha)}\}_{\ell=1}^{N-1}$ of associated spacings. Next, we compute the $\ell$-th mean spacing ${\Delta_\ell = M^{-1}\sum_{\alpha=1}^{M} \tilde{s}_\ell^{(\alpha)}}$ and construct a set of unfolded spacings ${\{ s_\ell^{(\alpha)} = \tilde{s}_\ell^{(\alpha)}/\Delta_\ell \}_{\ell=1}^{N-1}}$ for each sample $\alpha$.

In order to minimize potential fluctuations in numerically evaluated autocovariances, we perform both the (running) energy averaging within each sample,
\begin{eqnarray}\label{num-01}
    \delta {\mathcal I}_k^{(\alpha)} (N) = \frac{1}{N-1-k} \sum_{\ell=1}^{N-1-k} \left( s_\ell^{(\alpha)} s_{\ell+k}^{(\alpha)} - 1 \right),
\end{eqnarray}
and, on the top of it, the sample averaging
\begin{eqnarray}\label{num-02}
    \delta {\mathcal I}_k(N,M) = \frac{1}{M} \sum_{\alpha=1}^{M} \delta {\mathcal I}_k^{(\alpha)} (N).
\end{eqnarray}
So evaluated autocovariances of level spacings are plotted in Fig.~1. Due to statistical independence of random variables ${\{\delta {\mathcal I}_k^{(\alpha)} (N)\}_{\alpha=1}^M}$, the confidence interval at the $99\%$ level equals $\delta {\mathcal I}_k(N,M) \pm \sigma_k(N,M)$, where
\begin{eqnarray} \label{num-03}
    \sigma_k(N,M) = c_{0.99}\sqrt{\frac{{\rm var_\alpha}[\delta {\mathcal I}_k^{(\alpha)} (N)]}{M}}.
\end{eqnarray}
Here, ${\rm var_\alpha}[\cdots]$ is a sample variance, and $c_{0.99}\approx 2.58$. For all $2\le k \le 20$, the length $2\sigma_k(N,M)$ of the confidence interval turned out to be of the order $4 \times 10^{-6}$. Finally, we remark that the finite size effects~\cite{FM-2015}, which are expected to be of the order ${\mathcal O}(N^{-2})$, are small enough and should not affect our numerical results for the chosen sampling size.

\textit{\textbf{Acknowledgements.}}---This work was supported by the Israel Science Foundation through Grant No. 428/18. Some of the computations presented in this work were performed on the Hive computer cluster at the University of Haifa, which is partially funded through the Israel Science Foundation Grant No. 2155/15.


\begin{references}
\bibitem{Product-remark} We do not discuss an important case of the product statistics.
\bibitem{Rem-conventions} Throughout the Letter, we formally set $\lambda_0=0$ and $\lambda_{-1}=\lambda_1$, see also Eq.~(\ref{theta-k-relation}).
\end{references}
\end{document}